\documentclass[aps,prl,superscriptaddress,notitlepage,amsmath,amssymb,nofootinbib,longbibliography]{revtex4-1}
\usepackage[utf8]{inputenc} 
\usepackage{graphicx}
\usepackage{url}
\usepackage[bookmarks, pagebackref=false]{hyperref}
\usepackage[usenames,dvipsnames]{xcolor}

\definecolor{bla}{HTML}{03396C}
\definecolor{blaa}{HTML}{005B96}
\definecolor{blaaa}{HTML}{6497B1}

\hypersetup{
  colorlinks, 
  bookmarksopen, 
  bookmarksnumbered,
  citecolor=blaa, 		
  linkcolor=blaa,    	
  urlcolor=blaa,			
}

\usepackage{tikz}
\usetikzlibrary{intersections}
\usetikzlibrary{arrows,decorations.pathmorphing,backgrounds,positioning,fit,petri}
\usetikzlibrary{decorations.markings}
\usetikzlibrary{calc}
\usetikzlibrary{intersections,through,hobby}

\usepackage{natbib}

\newcommand{\MS}{\ensuremath{{\rm \overline{MS}}}}
\newcommand{\Tr}{\ensuremath{\mathrm{Tr}}}
\begin{document}

\title{
  \Large\color{bla} 
Four-Loop Gauge and Three-Loop Yukawa Beta Functions \\in a General Renormalizable Theory }

\author{Alexander {\sc Bednyakov}}\email{bednya@theor.jinr.ru} 
\affiliation{Bogoliubov Laboratory of Theoretical Physics, Joint Institute for Nuclear Research,
Dubna 141980, Russia}
\affiliation{P.N. Lebedev Physical Institute of the Russian Academy of Sciences, Moscow 119991, Russia}
\author{Andrey {\sc Pikelner}}\email{pikelner@theor.jinr.ru} 
\affiliation{Bogoliubov Laboratory of Theoretical Physics, Joint Institute for Nuclear Research,
Dubna 141980, Russia}

\begin{abstract}
  We present the beta functions of gauge and Yukawa couplings in general
  four-dimensional quantum field theory, at four and three loops, respectively.
  The essence of our approach is fixing unknown coefficients in the most general
  ansatz for beta functions by direct calculation in several simplified models. We
  apply our results to the Standard Model and its extension with an arbitrary
  number of Higgs doublets and provide expressions for all four-loop gauge
  couplings beta functions with matrix Yukawa interactions.
\end{abstract}
\maketitle

\section{Introduction} 
Renormalization group equations (RGE) find their numerous applications in many
physical problems formulated in the language of quantum field theory (QFT).
Being a convenient tool to improve perturbation theory results, RGE allow
one to obtain high-precision predictions for various quantities ranging from
critical exponents in the theory of critical phenomena to observables in the
Standard Model (SM) and its extensions.
In the early 1980s, Machacek and
Vaughn~\cite{Machacek:1983fi,Machacek:1983tz,Machacek:1984zw} presented their
two-loop RG functions in $\MS$ scheme for all dimensionless couplings in general
four-dimensional renormalizable QFT. Several misprints have been corrected
during the subsequent 35 years, and the classical result has been extended to
include RG functions for dimensionful
parameters~\cite{Luo:2002ti,Bednyakov:2018cmx,Schienbein:2018fsw,Sartore:2020pkk}.
The two-loop general expressions became highly demanded in studies of the SM
extensions. After the appearance of computer codes aimed to generate RGE for
user-specified models, two-loop RG studies have become \emph{de facto} standard in a
New Physics analysis.
After discovering the Higgs boson at the
LHC~\cite{Aad:2012tfa,Chatrchyan:2012ufa} and a burst of activity on the vacuum
stability problem (see, e.g., \cite{Bednyakov:2015sca}), it became clear that
three-loop RG
functions~\cite{Pickering:2001aq,Mihaila:2012fm,Bednyakov:2012rb,Bednyakov:2012en,Chetyrkin:2013wya,Bednyakov:2013eba}
can play an essential role in precision studies of the SM and its extensions.
Partial four-loop results in the
SM~\cite{Bednyakov:2015ooa,Zoller:2015tha,Chetyrkin:2016ruf} and three-loop beta
functions in the Two-Higgs-Doublet Model(THDM)~\cite{Herren:2017uxn} became
available during the past few years.
Recently, two major steps toward general high-order results have been made.
First of all, RG functions in general theories were represented by a linear
combination of independent tensor structures (TS) corresponding to contractions
of various indices. One can match these ``template'' expressions to known
results in specific models and extract model- independent
coefficients~\cite{Steudtner:2020tzo,Steudtner:2021fzs}. Second, new ideas based
on the so-called Weyl Consistency Conditions
(WCC)~\cite{Jack:2013sha,Antipin:2013sga,Poole:2019kcm} allow one to find
relations between known and unknown TS coefficients, thus, putting constraints
on to-be-computed numbers. In particular, WCC relate gauge, Yukawa, and
self-coupling beta functions computed at four, three, and two loops, respectively.
Let us also mention that WCC allows to resolve~\cite{Poole:2019txl} a well-known
issue with $\gamma_5$ ambiguity (see, e.g.,~\cite{Jegerlehner:2000dz}) in $\MS$
RG functions. Poole and Thomsen in Ref.~\cite{Poole:2019txl} use WCC to relate
the ambiguous four-loop terms in strong coupling beta
function~\cite{Bednyakov:2015ooa,Zoller:2015tha} to unambiguous three-loop
contributions to Yukawa beta functions~\cite{Chetyrkin:2012rz,Bednyakov:2012en}
and confirm the prescription advocated in Ref.~\cite{Bednyakov:2015ooa}. The
relation holds for any four-loop gauge beta function in a general QFT, and it
was immediately used for the gauge sector of the SM with diagonal Yukawa
matrices~\cite{Davies:2019onf}.
There is also an (unphysical) ambiguity in RG functions due to possible unitary
contributions~\cite{Fortin:2012hn,Bednyakov:2014pia,Herren:2017uxn} to field renormalization in models with flavor symmetries. In what follows, we assume a
natural choice of Hermitian anomalous dimension and refer to recent
work~\cite{Herren:2021yur} for more details.
In the current study, we present for the first time complete results for all TS
coefficients in gauge and Yukawa beta functions at four and three loops, respectively.
Partial results are already available in the
literature~\cite{Steudtner:2020tzo,Steudtner:2021fzs,Poole:2019kcm}, and we want
to emphasize the difference between our approach and methods used in previous
studies. In the latter, given WCC, the unknown beta-function coefficients are
fixed from \emph{known} results for specific (usually physical) models. One of
the most powerful constraints of this type is provided by the three-loop
calculation in THDM~\cite{Herren:2017uxn}.
On the contrary, we, for the first time, \emph{design simple toy models} with specific
gauge group structure, each giving constraints on different unknown
coefficients. We use diagrams for TS provided in Supplementary Materials of
Ref.~\cite{Poole:2019kcm} as a guide, but it is fair to say that we found the
required models by trial and error. We use TSs implemented in the prominent code
\texttt{RGBeta}~\cite{Thomsen:2021ncy} to speed up our investigation.
General four-dimensional QFT that covers most of possible phenomenological
applications can be written in the form~\cite{Poole:2019kcm}
\begin{align}
	\mathcal{L} & = - \frac{1}{4} G_{AB}^{-2} F^A_{\mu\nu} F^{B\mu\nu} 
                + \frac{1}{2} (D_\mu \phi)_a (D_\mu \phi)_a + \frac{i}{2} \Psi_i^T \gamma^\mu (D_\mu \Psi)_i - \frac{1}{2} \phi^a \Psi_i^T y_{aij} \Psi_j - \frac{1}{24} \lambda_{abcd}\phi_a \phi_b \phi_c \phi_d, 
                \label{eq:gen_lag}
\end{align}
where $A,B$ runs through all factors of the gauge group $\mathcal{G}$. The
coupling matrix $G_{AB}$ is symmetric and block diagonal with non-diagonal
entries corresponding to kinetic mixing between $U(1)$ factors present in
$\mathcal{G}$. The real scalars $\phi_a$ and Majorana fermions $\Psi_i$ belong
to some representation of $\mathcal{G}$. The Yukawa couplings are denoted by
$y_{aij}$ and are symmetric in fermion indices $ij$. The self-coupling
$\lambda_{abcd}$ is symmetric in all four scalar indices.
In this Letter, we consider four-loop $\beta^{(4)}_{AB}$ and three-loop
$\beta^{(3)}_{aij}$ contributions to gauge and Yukawa beta functions,
respectively. The latter are defined as
\begin{align}
	\beta_{AB} = \frac{ d G^2_{AB}}{d \ln \mu} & =  \frac{1}{2} \left[\sum\limits_l G^2_{AC} \frac{\beta^{(l)}_{CD}}{(4\pi)^{2l}} G^2_{DB} + (A \leftrightarrow B)\right], 
  & \beta_{aij} =\frac{ d y_{aij}}{d \ln \mu}  = &\frac{1}{2}
                                                   \left[
                                                   \sum\limits_l \frac{\beta^{(l)}_{aij}}{(4\pi)^{2l}}
                                                   + (i \leftrightarrow j)
                                                   \right].
\end{align}
and can be represented in terms of 202 and 308 TSs~\cite{Poole:2019kcm}, respectively,
\tikzset{
  di/.style={line width=1pt,draw=black, postaction={decorate},
    decoration={markings,mark=at position .65 with
      {\arrow[scale=1,draw=black,>=latex]{>}}}},
  fer/.style={line width=1pt,draw=bla},
  sca/.style={line width=1pt,draw=bla,densely dotted},
  glu/.style={line width=1pt,decorate, decoration={snake, segment length=2.2mm, amplitude=1mm}, draw=bla}
}
\begin{equation}
  \label{eq:betacoefdef}
  \beta_{AB}^{(4)} = \sum\limits_{n=1}^{202} \left(\mathfrak{g}_n^{(4)}\cdot
  \vcenter{\hbox{
      \begin{tikzpicture}
        \useasboundingbox (-1,-0.6) rectangle (1,0.6);
        \filldraw[fill=black!10!white, draw=black, line width=1pt] (0,0) circle (0.4cm);
        \draw[glu] (180:0.4) -- (180:1);
        \draw[glu] (0:0.4) -- (0:1);
        \draw (0,0) node[anchor=center] {$(n)$};
        \draw (-1,0.1) node[anchor=south] {\small $A$};
        \draw (1,0.1) node[anchor=south] {\small $B$};
      \end{tikzpicture}
    }}\right)
  , \quad
  \beta_{aij}^{(3)} = \sum\limits_{n=1}^{308} \left(\mathfrak{y}_n^{(3)}\cdot
  \vcenter{\hbox{
      \begin{tikzpicture}
        \useasboundingbox (-1,-0.8) rectangle (1,0.8);
        \draw[fer] (0,0) -- (200:1);
        \draw[fer] (0,0) -- (-20:1);
        \draw[sca] (90:0.4) -- (90:0.8);
        \filldraw[fill=black!10!white, draw=black, line width=1pt] (0,0) circle (0.4cm);
        \draw (0,0) node[anchor=center] {$(n)$};
        \draw (90:0.7) node[anchor=west] {$a$};
        \draw (200:1) node[anchor=south] {$i$};
        \draw (-20:1) node[anchor=south] {$j$};
      \end{tikzpicture}
    }} \right)
\end{equation}
with universal numerical coefficients $\mathfrak{g}^{(4)}_{n}$ and $\mathfrak{y}^{(3)}_{n}$. 
Our ultimate goal is to provide all these 202+308 numbers explicitly.

\section{Models}
We emphasize two different sources of constraints on beta-function coefficients.
One type is coming from direct calculations and fixes independently
$\mathfrak{g}^{(4)}$ and $\mathfrak{y}^{(3)}$. Another one is WCC providing
relations between these coefficients. Most of the first type constraints, which
one can obtain from SM and THDM, are already included in the analysis by Poole
and Thomsen \cite{Poole:2019kcm}. We extend these results by direct calculations
in toy $SU(n)$ gauge models described below. Calculation with arbitrary $n$
\cite{Cvitanovic:1976am} gives an additional handle on unknown coefficients.
Due to the nature of constraints from WCC, it is natural to consider Gauge and
Yukawa together. Since constraints from different sources are independent, the
combined system becomes overdetermined, and we have a large number of
equalities. The latter allows us not only to fix the TS coefficients but also to
make extensive crosschecks of the \texttt{RGBeta} code and the validity of our
results.

First of all, we consider an analog of scalar QCD with a gauge group
$\mathcal{G} = SU(n_1) \times SU(n_2)$, in which a fundamental scalar $\phi$ is
charged under both factors. The Lagrangian can be cast into the form
\begin{align}
	\mathcal{L}_{\mathbf{M1}} = -  
		\frac{1}{2 g_i^2} \Tr(F^{i}_{\mu\nu} F^{i}_{\mu\nu}) 
		+ (D_\mu \phi)^\dagger_{\alpha\rho} 
		  (D_\mu \phi)_{\alpha\rho} 
		- \frac{\lambda_1}{2}
		(\phi^\dagger _{\alpha\rho} \phi_{\alpha\rho})
		(\phi^\dagger _{\beta\sigma} \phi_{\beta\sigma})
		- \frac{\lambda_2}{2} 
		(\phi^\dagger_{\alpha\rho} \phi_{\alpha \sigma})
		(\phi^\dagger _{\beta\sigma} \phi_{\beta\rho}),
		\label{eq:lag_scalar_x2}
\end{align}
where $g_1$ and $g_2$ are gauge couplings. To carry out renormalization, we need
the self-interactions of scalars $\lambda_1$, $\lambda_2$ compatible with
$\mathcal{G}$. In eq.~\eqref{eq:lag_scalar_x2} we explicitly write group indices
$\alpha,\beta$, and $\rho,\sigma$ corresponding to fundamental representations
of $SU(n_1)$ and $SU(n_2)$, respectively.
The second model that we use is a gauge theory with single $SU(n)$. The spectrum
of the model consists of two fields in fundamental representation, a vectorlike
Dirac fermion $Q$ and a scalar $h$, and two singlet fields, a Weyl
spinor\footnote{In what follows, we use Dirac four-component spinors.} $u_R
\equiv P_R u$ and a scalar $s$:
\begin{align}
	\mathcal{L}_{\mathbf{M2}} & = - \frac{1}{4 g^2} F^a_{\mu\nu} F^a_{\mu\nu} + i \bar Q \gamma^\mu D_\mu Q + i \bar u_R \gamma^\mu \partial_\mu u_R + \frac{1}{2} (\partial_\mu s)^2 + |D_\mu h|^2 \nonumber\\
	& - y_s \bar Q Q s - y_u \left[ (\bar Q h) u_R + \mathrm{h.c.} \right]
	- \frac{\lambda_s s^4}{24} - \frac{\lambda_{sh}}{2} s^2 (h^\dagger h)
	- \frac{\lambda_{h}}{2} (h^\dagger h)^2. 
	\label{eq:lag_siglet_higgs}
\end{align}
Here $g$, $y_s$, and $y_u$ are gauge and Yukawa couplings of our interest, and
$\lambda_s$, $\lambda_{sh}$ and $\lambda_h$ are the required scalar
self-couplings.

We also study a gauge theory with $\mathcal{G} =
SU(n_1)\times SU(n_2) \times SU(n_3)$ describing interactions of a Dirac fermion
$\Psi$ in fundamental representation of each factor in $\mathcal{G}$ and three
adjoint scalars $\phi_i$, each charged only under one $SU(n_i)$:
\begin{align}
	\mathcal{L}_{\mathbf{M3}} & =  - \frac{1}{2 g_i^2} \Tr(F^{i}_{\mu\nu} F^{i}_{\mu\nu}) + \Tr[(D_\mu \phi_i) (D_\mu \phi_i)]
		+ i \bar \Psi \gamma_\mu (D_\mu \Psi)
		\nonumber\\
		    & 
		   -
		  y_i \left[
			\bar \Psi \phi_i \Psi + \mathrm{h.c.}
		\right]  
		    - \frac{\lambda_{ij}}{8} \Tr(\phi_i \phi_i) \Tr(\phi_j \phi_j)
		    - \frac{\lambda_{i}}{24} \Tr(\phi_i \phi_i \phi_i \phi_i),
	\label{eq:lag_sususu}
\end{align}
where summation over $i,j=1,3$ is assumed. The gauge and Yukawa couplings are
denoted by $g_i$ and $y_i$, respectively, and we have nine independent
self-couplings in the model, $\lambda_i$ and symmetric $\lambda_{ij}$.

Finally, we consider a $U(1)$ model with three Dirac fermions arranged as $\Psi = (\psi_1, \psi_2)$ and $\psi$.  They interact with charged ($h$) and neutral ($s$) Higgs bosons via matrix $(y_1)_{ij}$, vector $(y_2)_i$, $(y_3)_i$, and scalar $y_4$ Yukawa couplings: 
\begin{align}
	\mathcal{L}_{\mathbf{M4}} & =  - \frac{1}{2 g^2} (F_{\mu\nu} F_{\mu\nu}) + |D_\mu h|^2 + \frac{1}{2} (\partial_\mu s)^2 + i \bar \Psi_i \gamma_\mu (D_\mu \Psi_i) + i \bar \psi \gamma_\mu (D_\mu \psi) 
		\nonumber\\
		    & 
		    - \left[ (y_1)_{ij} s \bar \Psi_i P_R \Psi_j +
(y_2)_i h \bar \Psi_i P_R \psi +  (y_3)_i h^* \bar \psi P_R \Psi_i + y_4 s \bar \psi P_R \psi 
+ \mathrm{h.c.}
		\right]  
	- \frac{\lambda_s s^4}{24} - \frac{\lambda_{sh}}{2} s^2 (h^\dagger h)
	- \frac{\lambda_{h}}{2} (h^\dagger h)^2. 
	\label{eq:lag_172713}
\end{align}
The $U(1)$ charges satisfy $Q_{h} + Q_{\psi} = Q_{\Psi}$, and the sums run over $i=1,2$.

This choice of models is also motivated by the fact that we can easily implement
them both in \texttt{RGBeta}~\cite{Thomsen:2021ncy} and
\texttt{DIANA}~\cite{Tentyukov:1999is}. We use the former to obtain the beta
functions in terms of unknown coefficients, while the latter allows us to
utilize our standard setup \cite{Bednyakov:2012rb,Bednyakov:2015ooa} and compute
required two- and three-point functions with
\texttt{FORCER}~\cite{Ruijl:2017cxj}.
To extract RG functions for gauge and Yukawa couplings, we need one-loop
renormalization of the self couplings. We again use \texttt{RGBeta} to generate
the necessary Z-factors.
\section{Fixing coefficients}

With explicit results of calculation in models \textbf{M1}
(\ref{eq:lag_scalar_x2}), \textbf{M2} (\ref{eq:lag_siglet_higgs}), 
\textbf{M3} (\ref{eq:lag_sususu}) and \textbf{M4} (\ref{eq:lag_172713}) at hand, we are in position to apply all the
collected constraints and fix all beta-function coefficients. We summarize our
procedure in the Table \ref{tab:cFixNum}, where we show how the number of
unknowns $u_\mathfrak{g}$ and $u_\mathfrak{y}$ reduces after sequential
application of available constraints. We start with WCC connecting
$\mathfrak{g}^{(4)}$ and $\mathfrak{y}^{(3)}$ and interpret them as constraints
on gauge beta-function coefficients. Applying further constraints, we obtain new
relations (we denote the corresponding number by $\mathbf{n}$), and also a set
of identities (the corresponding number is given by ${c}$) for cross-checking.

\begin{table}[h]
  \centering
  \begin{tabular}{|l||c|c||c|c|}
	  \hline
    Type of the beta function & \multicolumn{2}{c||}{Gauge} & \multicolumn{2}{c|}{Yukawa}\\\hline
    Number of equations and unknowns & $r= \mathbf{n} + c$ & $u_\mathfrak{g}$ & $r = \mathbf{n} + c$ & $u_{\mathfrak{y}}$\\\hline\hline
    Initial number of unknown coefficients & - & 202 & - & 308\\\hline
    Weyl Consistency Conditions & \textbf{128}+0 & 74 & \textbf{133}+0 & 175 \\\hline
    Four-loop SM gauge beta functions &  \textbf{63}+84  & 11 &\multicolumn{2}{c|}{-}   \\\hline
    Three-loop matrix Yukawa beta functions in the SM  &  \multicolumn{2}{c||}{-}& \textbf{128}+17 & 47    \\\hline
    Three-loop matrix Yukawa beta functions in THDM &\multicolumn{2}{c||}{-} & \textbf{33}+213 & 14 \\\hline
    Four-loop QCD beta function for general group  & \textbf{2}+11 & 9 &\multicolumn{2}{c|}{-}\\\hline\hline
    $SU(n_1) \times SU(n_2)$ gauge theory \eqref{eq:lag_scalar_x2} 
    \hfill (\bf{M1}) & \textbf{5}+25 & 4 &\multicolumn{2}{c|}{-}\\\hline
    $SU(n)$ gauge theory \eqref{eq:lag_siglet_higgs} 
    \hfill (\bf{M2}) & \textbf{2}+55 & 2 & \textbf{4}+76 & 10\\\hline
    $SU(n_1)\times SU(n_2)\times SU(n_3)$ gauge theory \eqref{eq:lag_sususu} \hfill (\bf{M3}) & \multicolumn{2}{c||}{-} & \textbf{9}+89 & 1 \\\hline
    $U(1)$ gauge theory \eqref{eq:lag_172713} \hfill (\bf{M4}) & \multicolumn{2}{c||}{-} & \textbf{1}+199 & 0 \\\hline
    Constraints from symmetric  $T_{IJ}$   & \textbf{2}+8 & 0 & \multicolumn{2}{c|}{-}\\\hline\hline
    Final number of unknowns & & 0  & & 0\\
    \hline
  \end{tabular}
  \caption{Reduction of the number of unknown coefficients
    $u_\mathfrak{g}$ and $u_{\mathfrak{y}}$ after sequential application of
    constraints. Here $r = \mathbf{n} + c$ is the rank of the system without any previous
	  constraints included, $\mathbf{n}$ corresponds to new independent
    constraints, and $c$ relations are automatically satisfied due to previous steps.
  }
  \label{tab:cFixNum}
\end{table}

After matching template expressions with our toy models we are left with two
unknowns in the gauge sector. 
In Ref.~\cite{Poole:2019kcm} the authors conjectured that $T_{IJ}$ tensor
entering WCC can be symmetric. This provides additional 10 constraints. We use 2
of them to constrain the remaining coefficients. The other 8 equations become
identities and, thus, verify the assumption on $T_{IJ}$.
It is worth noting that we use models \textbf{M1}-\textbf{M3} to constrain all TSs  but the difference $\mathfrak{y}^{(3)}_{172} - \mathfrak{y}^{(3)}_{173} \equiv 2 \delta$ in Yukawa beta function. 
To deal with $\delta$, we developed  a special model \textbf{M4}, which explicitly confirms our initial guess that $\delta=0$.

With this procedure we fix all coefficients in gauge and Yukawa beta functions at four and three loops, respectively. 

\section{Results and discussion}
We combine the constraints from WCC~\cite{Jack:2013sha} and our explicit
computations in toy models to fix all the coefficients in the ansatz for the
4-3-2 ordering given in Ref.~\cite{Poole:2019kcm}. As an application of our
general expressions, we derive all four-loop gauge beta functions in the SM
extension with the arbitrary number $n_d$ of Higgs doublets (NHDM). We keep
matrix Yukawa couplings, and by setting $n_d=1$ or $n_d=2$, we obtain the SM and
THDM result. Direct computation in such a scenario would be very complicated.
Let us return to the ambiguities in theories possessing a flavor symmetry. In
Refs.~\cite{Fortin:2012hn,Poole:2019kcm,Herren:2021yur}, for the coupling
$y_{aij}$ a flavor-improved, i.e unambiguous, version $B_{aij}$ of the Yukawa
beta function $\beta_{aij}$ is introduced
\begin{align}
	B^{(3)}_{aij} = \beta^{(3)}_{aij} - S^{(3)}_{ik} y_{akj} - S^{(3)}_{jk} y_{aik} - S^{(3)}_{ab} y_{bij}.
	\label{eq:beta_flavour_improved_1}
\end{align}
Here $S^{(3)}_{ij}$ and $S^{(3)}_{ab}$ are three-loop quantities that can be
represented as linear combinations of antisymmetric two-point TS for fermions
and scalars, respectively \cite{Fortin:2012hn,Poole:2019kcm}. There are six
coefficients $\mathfrak{f}^{(3)}_{1-6}$ entering $S^{(3)}_{ij}$ and three
coefficients $\mathfrak{s}^{(3)}_{1-3}$ entering $S^{(3)}_{ab}$. All but one
($\mathfrak{f}^{(3)}_{3}$) numerical coefficients are predicted from
WCC\footnote{See Supplementary Material of Ref.~\cite{Poole:2019kcm}.} in terms of
$\mathfrak{g}^{(4)}_i$ and $\mathfrak{y}^{(3)}_i$ computed in this Letter. The
authors of Ref.~\cite{Herren:2021yur} calculated $\mathfrak{f}^{(3)}_{4} = -3/8
$ and $\mathfrak{f}^{(3)}_5=-5/16$. Given our results, we provide full set of three-loop corrections to $S_{ij}$ and $S_{ab}$:
\begin{align}
	\mathfrak{f}^{(3)}_1 & = 0, 
  & \mathfrak{f}^{(3)}_2 & = \frac{29}{8} - 3 \zeta_3, 
  & \mathfrak{f}^{(3)}_3 & = \frac{21}{8} - 3 \zeta_3,
  & \mathfrak{f}^{(3)}_4 & = -\frac{3}{8}, 
  & \mathfrak{f}^{(3)}_5 & = -\frac{5}{16}, 
  & \mathfrak{f}^{(3)}_6 & = -\frac{7}{16}. 
\label{eq:S_fermionic}
\end{align}
\begin{align}
	\mathfrak{s}^{(3)}_1 & = \frac{7}{2} - 6 \zeta_3, 
  & \mathfrak{s}^{(3)}_2 & = \frac{5}{8},
  & \mathfrak{s}^{(3)}_3 & = - \frac{3}{4}.
                           \label{eq:S_scalar}
\end{align}
These coefficients can be tested by direct calculations along the lines of
Refs.~\cite{Fortin:2012hn,Herren:2021yur}.

\section{Conclusion}
The calculation of four-loop Gauge and three-loop Yukawa beta functions
performed in this Letter complement recent six-loop results in general pure
scalar theory \cite{Bednyakov:2021clp}, and represents the most advanced
achievement in this field. The obtained TS coefficients can be incorporated into
modern computer codes, giving access to a new precision level for model
building. The dummy-field method (see, e.g., Ref.~ \cite{Staub:2013tta}) applied
to our results provides us with scale dependence of such important quantities as
fermion mass matrices.
We make all our results, including the TS coefficients and four-loop
gauge-coupling beta functions in the SM, THDM, and NHDM, available as
Supplementary Material. We also provide a modified version of the RGBeta package
with all our findings included~\cite{rgbeta}.
\acknowledgments
We thank A.E. Thomsen for correspondence regarding \texttt{RGBeta} and
Ref.~\cite{Poole:2019kcm}. We are grateful to the Joint Institute for Nuclear
Research for letting us use their supercomputer “Govorun.” The work of A.P. is
supported by the Foundation for the Advancement of Theoretical Physics and
Mathematics ``BASIS.'' The work of A.B. is supported by the Grant of the Russian
Federation Government, Agreement No. 14.W03.31.0026.
\bibliography{beta432.bib}
\end{document}